\documentclass{ws-ijmpcs}

\begin{document}

%
%

\title{Development of a high-sensitivity torsion balance to investigate the thermal Casimir force}

\author{Todd Graveson, Charles Rackson, Woo-Joong Kim\footnote{kimw@seattleu.edu}}
\address{Department of Physics, Seattle University\\ 901 12th Avenue, Seattle, WA 98122, USA}

\maketitle

\begin{abstract}
We report development of a high-sensitivity torsion balance to measure the thermal Casimir force. Special emphasis is placed on experimental investigations of a possible surface electric force originating from surface patch potentials that have been recently noticed by several experimental groups. By gaining a proper understanding of the actual contribution of the surface electric force in real materials, we aim to undertake precision force measurements to resolve the Casimir force at finite temperature in real metals, as well as in other semiconducting materials, such as graphene.
\end{abstract}

\ccode{PACS numbers: 12.20.Fv, 07.10.Lw, 68.65.Pq, 68.47.De}

\section{Introduction}	

The Casimir force has attracted much attention during the last decade. This is due to the simple, but elegant approach originally proposed by H. B. Casimir\cite{Casimir} to predict a macroscopic electrodynamic force between a pair of metallic plates. Starting from the landmark demonstration of the Casimir force using a torsion balance by S. K. Lamoreaux\cite{Steve} in 1997, numerous measurements employing different experimental techniques verifying the existence of the force have ensued\cite{Mosterev,Adrian}. Most notably, it has been shown that the Casimir force becomes repulsive by tailoring the dielectric properties of the employed materials\cite{Note1} and their respective geometries\cite{Johnson}, opening up new possibilities\cite{demanPRL,ChanPRL,ChanPRL2,MohideenPRL} for nano-electromechanical systems (NEMS). Particularly, the repulsive Casimir force has been proposed to mitigate friction between closely spaced mechanical parts in a NEMS device\cite{Kenneth}. 

Incidentally, a number of experimental groups investigating the Casimir force\cite{PRARC,Proceed,deman,Yaleprl} as well as experimental colleagues from other research areas in physics have indicated the pervasive presence of the surface electric forces that are inevitably present on the surface of real, metallic conductors\cite{myreview,MITion,Naji}. Examples include: fundamental studies of friction and force originating from van der Waals' dispersion interactions\cite{Burnham,Stipe}, large scale projects like Gravity Probe-B (GPB), Laser Interferometer Space Antenna (LISA) and Laser Interferometer Gravitational wave Observatory (LIGO)\cite{GPB,Speake}, and ion trap experiments for quantum computing\cite{Berkeley}, just to name a few. Very commonly---yet repeatedly---reported in all of the aforementioned experiments is the difficulty of quantifying systematic noise associated with the surface patch potentials that exist on the surface of the usually metallic test bodies.

Various models have been proposed in an attempt to capture the essential physics underlying a particular experiment in the midst of the surface electric noises, but no definitive consensus has been reached so far, regarding the extent to which the surface interactions caused by patch potentials for a given experiment are actually present. As a primary example, the observation of the thermal Casimir force reported in Ref. (\refcite{Yalenature}) has been recently questioned\cite{Moste,MostePRB}, mainly because of the presence of a large surface electric force. This large surface force had to be subtracted from the total measured force in order to confirm a particular theoretical model (e.g. the Drude model). A similar approach had been taken earlier when Burnham \textit{et al}.\cite{Burnham} observed an unusually large contribution of long-range electric interactions, which also had to be distinguished from the expected van der Waals interaction between their metallic test bodies. In the friction measurements reported by Stipe \textit{et al.}\cite{Stipe}, the surface electric effect is attributed to be the largest systematic noise, thereby setting a fundamental limit to modern precision measurements involving two closely spaced bodies. Intense experimental activities to understand the origin of the surface electric effect\cite{Bridgman,Dowling,Rose} began as early as 1919, when the metallic work function difference is shown to exhibit a noticeable variation under different experimental conditions. Surprisingly, there have not yet been tangible efforts to create a cohesive---and collaborative---experimental program to effectively deal with systematic noises associated with the recurrent surface electric effect. 

Here, we report on the development of a high-sensitivity torsion balance to investigate various surface interactions in common metals originating from both electric (patch potentials) and quantum-mechanical (Casimir) effects. Emphasis will be placed on dedicated force measurements under different experimental conditions to provide reliable measures of the actual contribution of the Casimir force, thereby yielding meaningful confirmations of the prediction of the Lifshitz's theory combined with a particular model\cite{Sernel,Milton1,Milton2}. With improved sensitivity, we hope to extend our force measurements beyond metallic samples  to a system consisting of graphene, which appears extremely promising in exploring the finite-temperature Casimir effect\cite{Bordag,Sernelgrap,Valery,Svet}.  

\section{Instrument}
Our apparatus consists of a torsion balance hung by a 20-cm long tungsten fiber ($D$=76 $\mu$m) in a vacuum chamber (bell jar), and employs an optical lever technique\cite{Sasaki1,Sasaki2} to measure a force between ``Casimir plates'', as shown in Fig. 1. On one side of the torsion balance is a flat plate facing a spherical plate of radius of curvature $R$ attached to a closed-loop piezoelectric transducer (P841.10, Physik Instrumente), whose position is additionally controlled by coarse motions of a three-axis stage (T25, Thorlabs). In fact, this mechanical stage alone has displacement resolution of 8 nm and could be used to replace the PZT. To investigate the possible effects of the background electric noise originating from the materials of the torsion balance itself\footnote{This was suggested by Dr. Riccardo Decca at the recent QFEXT conference: Oxidation on the surface of materials made of Al may result in a large electric noise, especially when the employed test plates (Casimir plates) are made of another type of metal.}, we have constructed two identical balances, one being made of Al and the other one electroplated with Au.  

To avoid difficulties maintaining the parallelism of the Casimir plates, we have chosen a sphere-plane geometry. One of  our experimental objectives is to make a critical examination on the linear relationship between the measured force and the radius of curvature $R$ of a given sphere. As such, we have prepared three different materials, each with varying $R$-values, as summarized in Table 1. In Refs. (\refcite{MostePRB,Decca}), it is speculated that the cm-sized macroscopic plates tend to give rise to anomalous behaviors during electrostatic calibrations (e.g., the distance-dependent CPD and deviations from the expected electrostatic power law\cite{PRARC,Reply}), because of their intrinsic surface imperfections. Utilizing different spheres of varying sizes in a single apparatus that is well-calibrated will eventually help us better understand the extent to which the use of large plates (cm-sized) affects the actual force measurements. Each of the plates is coated with a sacrificial layer of Cr followed by 1000 {\AA} of Au by thermal evaporation. Samples containing single or multil-layers of  graphene have also been prepared on a Si wafer\footnote{We are currently collaborating with Dr. Daniil Stolyarov at Graphene Laboratories Inc. to fabricate the samples.}. Large area synthesis of uniform graphene layers is now possible, and cm-sized graphene films are routinely produced in large quantities\cite{Rodney}.          

When a force is applied between the Casimir plates, the torsion pendulum experiences a torque and its rotation is detected by a laser deflection collected on a quadrant photodiode. The deflection is then converted into error signals $\delta V$ via a Proportional-Integral-Differential (PID) controller and is sent to the feedback plates located on the other side of the balance, to keep the balance angle fixed. These deflection signals are obtained from the difference between the horizontal readings of the quadrant photodetector with an estimated sensitivity of 0.5 mV/$\mu$rad. 

A high-vacuum of order $10^{-6}$ mbar will be maintained using roughing and diffusion pumps. The torsion pendulum and the vacuum chamber sit on a vibration isolation table capable of pneumatic self-leveling with an air-compressor. Vertical signals from the quadrant photodiode detector are used to monitor seismic and tilt motions.          

\section{Projected sensitivity}
The feedback (correction) signal $\delta V$ is proportional to the total force experienced by the Casimir plates:
\begin{equation}
\delta V(d,V)\propto F_{\rm{total}}(d,V)=F_{\rm{elect}}(d,V)+F_{\rm{Cas}}(d),
\label{force}
\end{equation}
where $F_{\rm{elect}}(d,V)=\pi R \epsilon_0(V-V_0)^2/d$. $F_{\rm{Cas}}(d)$ represents the Casimir force, which is purely distance-dependent. Electrostatic calibration is conducted by applying a set of external electric voltages $V$, thereby minimizing the electric force $F_{\rm{elect}}$ at $V=V_0$ at different distances. Additionally, it enables the characterization of the calibration factor $\beta$, converting the units of the feedback signals (V) into the actual units of force (N). Note that the PZT records only the relative distance $d_r$, and therefore, a point of contact ($d_0$) must be obtained in order to measure absolute gap distances: $d\equiv d_0-d_r$. Accordingly, the distance dependence of $V_0$, defined as a contact potential difference (CPD), is also investigated during the electrostatic calibration. Detailed calibration procedures are described in Refs. (\refcite{PRARC,demanJVST}). 

Since the publication of Ref.(\refcite{PRARC}) in 2008 reporting an anomalous behavior related to the distance-varying contact potential difference, a number of experimental groups have carefully undertaken electrostatic calibrations during their Casimir force experiments---and many of them have repeatedly confirmed the existence of the distance-dependent CPD\cite{demanJVST,Toricel,ToricelEPL,Qun}. Moreover, in addition to the distance-varying minimizing potential $V_0$,  a possible influence of electric patch potentials has been suggested, which could persist at the surface of the metallic objects employed during the Casimir force measurements, even at the electrically minimized conditions\cite{Yaleprl,Yalenature,YalePRA,Ryan,Mos}. As of writing this manuscript, J. Laurent \textit{et al.}\cite{Joel} have just reported the observation of the distance-varying contact potential {\it and} the presence of a long-range patch force in Au-Au and Au-Si samples at low temperature (4.2 K). Clearly, despite careful nullification of the electric force $F_{\rm{elect}}$ at $V=V_0$ in Eq. 1, an electric force still exists in the form of patch force and must be taken into account:
\begin{equation}
F_{\rm{total}}(d,V_0)=F_{\rm{Cas}}(d)+F_{\rm{patch}}(d),
\label{patch}
\end{equation}
where we define $F_{\rm{elect}}(d,V_0)\equiv F_{\rm{patch}}(d)$. This additional electric patch force must be now distinguished from the pure Casimir force, both of which are distance-dependent. Much of the debate contained in Refs.(\refcite{Yaleprl,Yalenature,Moste,MostePRB}) concerns the accuracy of the exact quantification of $F_{\rm{patch}}(d)$ that needs to be subtracted from a total measured force to reveal the actual thermal contribution of the Casimir force. We believe that this particular issue remains to be resolved, and it is our hope that subsequent measurements enabled by our apparatus  could make a meaningful contribution to the ongoing debate.   

To estimate the sensitivity of our torsion pendulum, we first consider fluctuations due to instability of a positioning device. Note that the PZT employed in our apparatus has position accuracy of 0.2 nm and operates in a closed-loop. Based on this, the minimum electric and Casimir forces to be resolved at 1 $\mu$m is 0.025 pN and 0.038 pN, respectively, assuming $R$=15.5 cm and $V_{\rm{patch}}$=5 mV. Even if we take a scenario in which the position accuracy worsens by an order of magnitude, the force sensitivity at 1 $\mu$m is still less than 1 pN. 

Realistically, the overall force sensitivity is limited by angular resolution (1 $\mu$rad/mV) of the optical lever scheme employed in our measurements, similarly reported in Ref. (\refcite{Sasaki2}). The minimum angle of deflection (for $\delta V=0.1$ mV) is approximately $\delta\theta_{\rm{min}}$=0.1 $\mu$rad. Based on the torsion angular spring constant\cite{Torg} $\alpha=2.96\times10^{-6}$ N m/rad for the torsion modulus $Z=1.8\times10^{11}$ N/${\rm m}^2$, we have a force resolution of less than  3 pN. Note that thermal noise based on the fluctuation-dissipation theorem, associated with the angular motion $\delta\theta_{\rm{a}}=\sqrt{k_bT/\alpha}=3.8\times10^{-8}$ rad, which is still an order of magnitude less than the angular resolution set by our detector at room temperature ($T$=300 K and $k_b$ the Boltzmann constant). We believe that this is the current limitation of our force sensitivity. By contrast, thermal noise related to the swinging motion of our torsion under the influence of gravitational acceleration $g$ with $m=97.3$ g is much less $\delta\theta_{\rm{s}}=\sqrt{k_bT/mgl}=1.5\times10^{-10}$ rad and therefore may be neglected. With the projected force sensitivity of 3 pN and in the absence of (or with the exact quantification available) the residual electric force $F_{\rm{patch}}(d)$, the thermal contribution in the Casimir force in metals can be readily resolved at a distance as large as 5 $\mu$m.    

\section{Current status}
Our apparatus (shown in Fig. 2) is currently located in Bannan Science Building at Seattle University in Seattle, WA (USA). The base pressure of $10^{-6}$ mbar has been achieved, and electric calibrations are currently underway. 

We have tested feedback stability in an ambient pressure without implementing any active vibration control, as shown in Fig. 3. Given the feedback voltage fluctuation on the order of 200 mV, our apparatus has already reached a force resolution of $\mu$N. A further control using active vibration isolation along with a high-vacuum operation should bring the feedback signal down to 1 mV or less, translating into a pN force sensitivity. 

The entire apparatus occupies an area of less than 15 ft$^2$ and is easy to transport. Final experimentation including calibration and force measurements will be performed in an isolated, noise-free laboratory room with active humidity and temperature controls. Computerized data acquisition and control will also be implemented during the final measurement stage.   

We have also developed a Michelson's interferometer to independently calibrate our PZT. The interferometer enables us to confirm the conversion factor of the employed PZT- from the applied voltages (V$_{\rm{PZT}}$) to actual displacements in meters- by utilizing the emission wavelength of a He-Ne laser (See Fig. 4). 

\section{Conclusions}
We have reported the development of a torsion pendulum aimed at investigating the thermal Casimir force, with special emphasis on various systematic issues surrounding the surface electric effect. Of foremost importance in our experimental objectives is a critical examination of the surface electric (residual) force, possibly caused by the presence of patch potentials at the surface of real materials. This surface electric force could {\em persist}, even if $F_{\rm{elect}}$ is made to be minimized during an electrostatic calibration. With the projected force sensitivity at the level of 3 pN, it is feasible to explore the thermal effect of the Casimir force, thereby distinguishing a particular theoretical model to be utilized in the Lifshitz theory. Looking forward, with the recent debate spurred by Refs.(\refcite{Yalenature,Moste,MostePRB}) along with the similar findings reported by other experimental groups outside the Casimir community, we emphasize the need for creating extensive collaborations\cite{myreview} with unified efforts to attack one of the pervasive sources of systematic errors stemming from surface patch potentials in precision force measurements. 

\section*{Acknowledgments}
Part of this work was conducted at the University of Washington NanoTech User Facility (NTUF), a member of the NSF National Nanotechnology Infrastructure Network (NNIN). This work was also supported by the M. J. Murdock Charitable Trust and by Research Corporation through Single-Investigator Cottrell College Science Award (SI-CCSA). We also thank S. K. Lamoreaux for his many insightful comments on the construction of the torsion balance.

\newpage
\begin{figure}[pb]
\centerline{\psfig{file=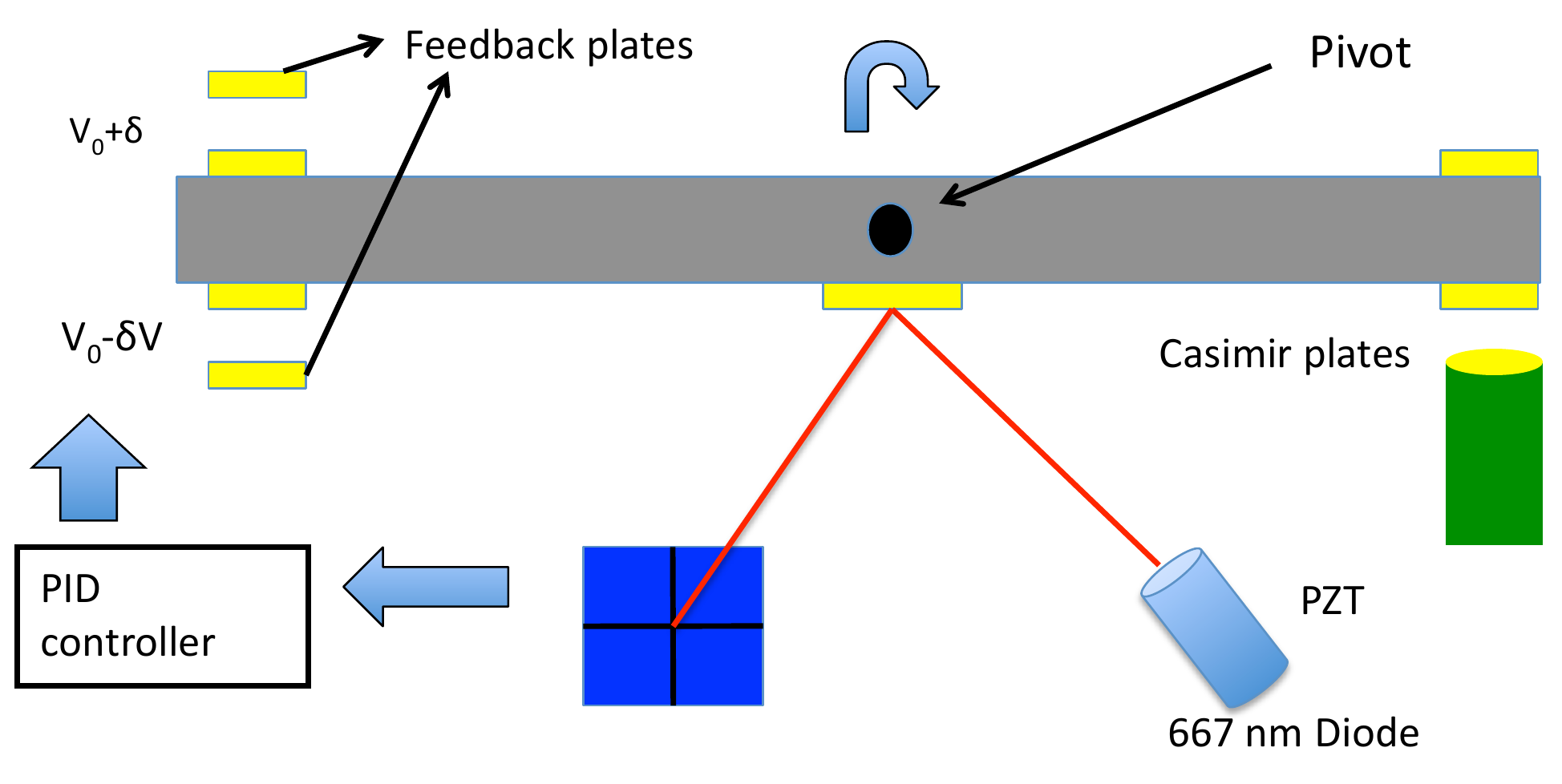,width=10cm}}
\vspace*{8pt}
\caption{Schematic of basic operations of our torsion balance experiment: An attractive force between the two ``Casimir plates'' exerts a torque in the clockwise direction, causing a deflection of the reflected laser beam off a mirror surface attached on the balance. The degree of deflection is converted into an error signal, which is applied to the ``feedback plates'' to restore the original angle (null measurements). A PZT attached to one of the Casimir plates varies the gap distance separating the two plates.}
\end{figure}

\begin{table}[pb]
\tbl{Available samples of different radii of curvature ($R$)}
{\begin{tabular}{@{}ccc@{}} \toprule
BK quartz (cm sized) & Diode lens (mm sized) & Polystyrene beads ($\mu$m sized) \\
& &\\
10.3 cm & 0.55 mm & 45 $\mu$m \\
15.5 cm & 1.10 mm & 110 $\mu$m  \\
30.9 cm  & 1.65 mm & 380 $\mu$m  \\
154.5 cm & 2.75 mm & 600 $\mu$m \\ \botrule
\end{tabular} \label{ta1}}
\end{table}

\begin{figure}[pb]
\centerline{\psfig{file=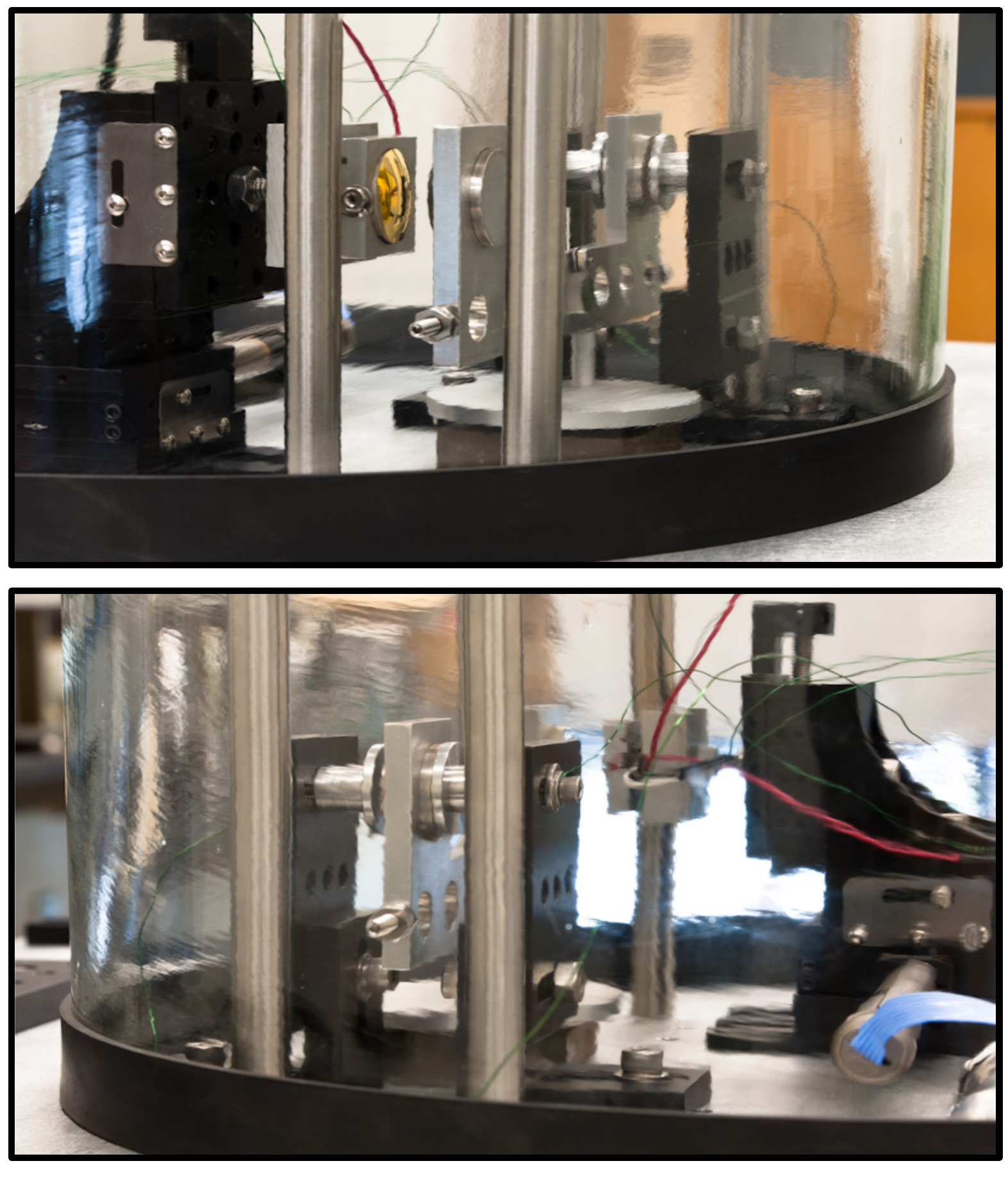,width=9cm}}
\vspace*{8pt}
\caption{Displayed are the Casimir plates (top) and the feedback plates (bottom). The torsion balance depicted in the figure is made of Al, but we also have an identical balance electroplated with Au. One of the Casimir plates is thermally insulated with a thermoelectric cooler (TEC) and a thermistor. The temperature control unit has a resolution of 0.5 K and varies over the range $T$=200-300 K.}
\end{figure}

\begin{figure}[pb]
\centerline{\psfig{file=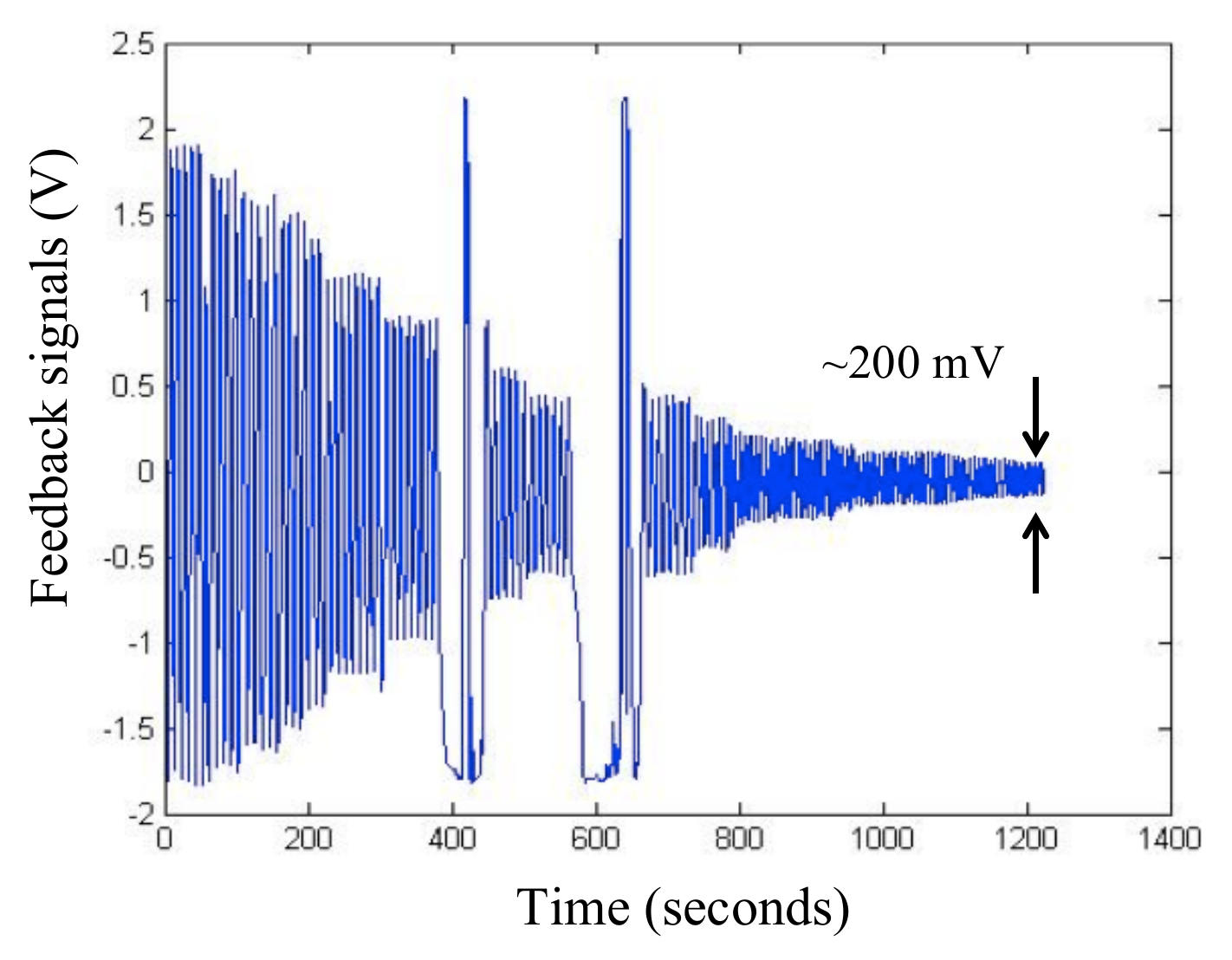,width=9cm}}
\vspace*{8pt}
\caption{Test of negative feedback circuitry implemented for null measurements. A stable, feedback signal down to 200 mV has been achieved in ambient pressure without any active vibration control. A further stabilization on the order of 1 mV or less is to be expected in the future. Occasional spikes seen on the plot are due to accidental disturbance.}
\end{figure}

\begin{figure}[pb]
\centerline{\psfig{file=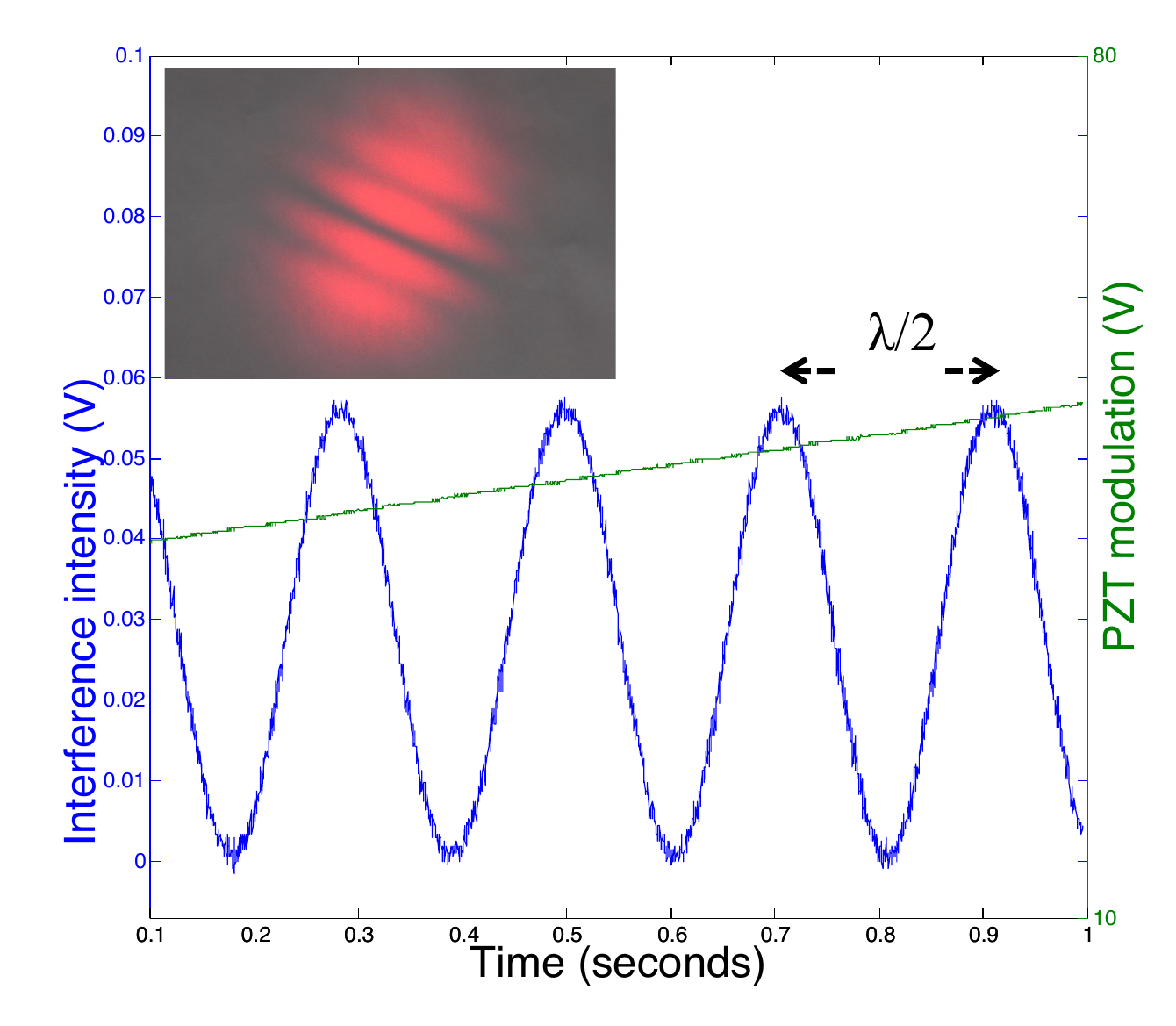,width=9cm}}
\vspace*{8pt}
\caption{Implementation of Michelson's interferometer to calibrate a PZT. Fringe is clearly seen in the inset, while a laser beam (He-Ne laser) experiences interference whose intensity is periodically modulated over a visibility larger than 90\%. Periodicity of $\lambda/2$, where $\lambda=632.8$ nm is the laser wavelength, enables a direct calibration of the PZT's applied voltage to the actual position displacement.}
\end{figure}


\begin{thebibliography}{000}    
\bibitem{Casimir} H. B. Casimir, {\it Proc. K. Ned. Akad. Wet. B} {\bf 51}, 793 (1948).
\bibitem{Steve} S. K. Lamoreaux, {\it Phys. Rev. Lett.} {\bf 78}, 5 (1997).
\bibitem{Mosterev} M. Bordag, G. L. Klimchitskaya and V. M. Mostepanenko,{\it Advances in the Casimir effect} (Oxford University Press, Oxford, 2009).
\bibitem{Adrian} V. A. Parsegian, {\it Van der Waals forces} (Cambridge University Press, Cambridge, 2006).
\bibitem{Lifshitz} E. M. Lifshitz {\it J. Exp. Theor. Phys} {\bf 2}, 73 (1956).
\bibitem{Rep1} A. Milling, P. Mulvaney and I. Larson {\it J. Colloid Interface Sci.} {\bf 180}, 460 (1996).
\bibitem{Rep2} A. Meurk, P. F. Luckham and L. Bergstrom, {\it Langmuir} {\bf13}, 3896 (1997).
\bibitem{Rep3} S. Lee and W. M. Sigmund, {\it J. Colloid Interface Sci.} {\bf 243}, 365 (2001).
\bibitem{Note1} The repulsive dispersion force is predicted in the context of the Lifshitz theory\cite{Lifshitz} and experimentally investigated by several groups\cite{Rep1,Rep2,Rep3}.  
\bibitem{Johnson} A. W. Rodriguez, F. Capasso and S. G. Johnson, {\it Nature Photon.} {\bf 5}, 211 (2011).
\bibitem{demanPRL} S. de Man, K. Haek, R. J. Wijngaarden and D. Iannuzzi, {\it Phys. Rev. Lett.} {\bf 103}, 040402 (2009).
\bibitem{ChanPRL} H. B. Chan, Y. Bao, J. Zou, R. A. Cirelli, F. Klemens, W. M. Mansfield and C. S. Pai, {\it Phys. Rev. Lett.} {\bf 101}, 030401 (2008).
\bibitem{ChanPRL2} Y. Bao, R. Gu{\'e}rout, J. Lussange, A. Lambrecht, R. A. Cirelli, F. Klemens, W. M. Mansfield, C. S. Pai and H. B. Chan, {\it Phys. Rev. Lett.} {\bf 105}, 250402 (2010).
\bibitem{MohideenPRL} C. C. Chang, A. A. Banishev, G. L. Klimchitskaya, V. M. Mostepanenko and U. Mohideen, {\it Phys. Rev. Lett.} {\bf 107}, 090403 (2011).
\bibitem{Kenneth} O. Kenneth, I. Klich, A. Mann and M. Revzen, {\it Phys. Rev. Lett.} {\bf89}, 033001 (2002).

\bibitem{PRARC} W. J. Kim, M. Brown-Hayes, D. A. R. Dalvit, J. H. Brownell and R. Onofrio, {\it Phys. Rev. A} {\bf 78}, 020101(R) (2008).
\bibitem{Proceed} W. J. Kim, M. Brown-Hayes, D. A. R. Dalvit, J. H. Brownell and R. Onofrio, {\it J. Phys. Conf. Ser} {\bf 161}, 012004 (2009).
\bibitem{deman} S. de Man, K. Haek and D. Iannuzzi, {\it Phys. Rev. A} {\bf 79}, 024102 (2009).
\bibitem{Yaleprl} W. J. Kim, A. O. Sushkov,  D. A. R. Dalvit  and S. K. Lamoreaux, {\it Phys. Rev. Lett.} {\bf 103}, 060401 (2009).

\bibitem{Burnham} N. Burnham, R. J. Colton and H. M. Pollock, {\it Phys. Rev. Lett.} {\bf 69}, 144 (1992).
\bibitem{Stipe} B. C. Stipe, H. J. Mamin, T. D. Stowe, T. W. Kenny and D. Rugar, {\it Phys. Rev. Lett.} {\bf 87}, 096801 (2001).
\bibitem{GPB} GPB Collab. (C. W. Everitt {\it et al}.),  {\it Phys. Rev. Lett.} {\bf 106}, 221101 (2011).
\bibitem{Speake} C. C. Speake and C. Trenkel, {\it Phys. Rev. Lett.} {\bf 90}, 160403 (2003).
\bibitem{Berkeley} N. Daniilidis and H. H{\"a}ffner, {\it Physics} {\bf 4}, 66 (2011).

\bibitem{myreview} W. J. Kim and U. D. Schwarz, {\it J. Vac. Sci. Technol. B} {\bf 28}, C4A1 (2010).
\bibitem{MITion} J. Labaziewicz, Y. Ge, D. R. Leibrandt, S. X. Wang, R. Shewmon and I. L. Chuang {\it Phys. Rev. Lett.} {\bf 101}, 180602 (2008).
\bibitem{Naji} A. Naji, D. S. Dean, J. Sarabadani, R. R. Horgan and R. Podgornik, {\it Phys. Rev. Lett.} {\bf 104}, 060601 (2010).

\bibitem{Yalenature} A. O. Sushkov, W. J. Kim, D. A. R. Dalvit and S. K. Lamoreaux, {\it Nature Phys.} {\bf 7}, 230 (2011).
\bibitem{Moste} G. L. Klimchitskaya, M. Bordag, E. Fischbach, D. E. Krause and V. M. Mostepanenko, {\it J. Mod. Phys. A} {\bf 26}, 3918 (2011).
\bibitem{MostePRB} V. B. Bezerra, G. L. Klimchitskaya, U. Mohideen, V. M. Mostepanenko and C. Romero, {\it Phys. Rev. B} {\bf 83}, 075417 (2011).

\bibitem{Bridgman} P. W. Bridgman, {\it Phys. Rev.} {\bf 14}, 306 (1919).
\bibitem{Dowling} P. H. Dowling, {\it Phys. Rev.} {\bf 31}, 244 (1928).
\bibitem{Rose} B. A. Rose, {\it Phys. Rev.} {\bf 44}, 585 (1933).

\bibitem{Bordag} M. Bordag, I. V. Fialkovsky, D. M. Gitman and D. V. Vassilevich,  {\it Phys. Rev. B} {\bf 80}, 245406 (2009).
\bibitem{Sernelgrap} Bo. E. Sernelius, {\it Europhys. Lett. } {\bf 95}, 57003 (2011).
\bibitem{Valery} I. V. Fialkovsky, V. N. Marachevsky and D. V. Vassilevich, {\it Phys. Rev. B} {\bf 84}, 035446 (2011).
\bibitem{Svet} V. Svetovoy, Z. Moktadir, M. Elwenspoek and H. Mizuta, {\it Europhys. Lett. } {\bf 96}, 14006 (2011).

\bibitem{Sernel} M. Bostr\"{o}m and B. E. Sernelius, {\it Phys. Rev. Lett.} {\bf 84}, 4757 (2000).
\bibitem{Milton1} J. S. H{\o}ye, I. Brevik, J. B. Aarseth and K. A. Milton, {\it Phys. Rev. E} {\bf 67}, 056116 (2003).
\bibitem{Milton2} I. Brevik,  J. B. Aarseth, J. S. H{\o}ye and K. A. Milton, {\it Phys. Rev. E} {\bf 71}, 056101 (2005)

\bibitem{Sasaki1} M. Masusa and M. Sasaki, {\it Phys. Rev. Lett.} {\bf 102}, 171101 (2009).
\bibitem{Sasaki2} M. Masusa, M. Sasaki and A Araya, {\it Class. Quantum Grav.} {\bf 24}, 3965 (2007).

\bibitem{Decca} R. S. Decca, E. Fischbach, G. L. Klimchitskaya, D. E. Krause, D. L{\'o}pez, U. Mohideen and V. M. Mostepanenko, {\it Phys. Rev. A} {\bf 79}, 026101 (2009).
\bibitem{Reply} W. J. Kim, M. Brown-Hayes, D. A. R. Dalvit, J. H. Brownell and R. Onofrio, {\it Phys. Rev. A} {\bf 79}, 026102 (2009).
\bibitem{Rodney} X. Li W. Cai, J. An, S. Kim, J. Nah, D. Yang, R. Pinter, A. Velamakanni, I. Jung, E, Tutuc, S. K. Banerjee, L. Colombo and R. S. Ruoff, {\it Science} {\bf 324}, 1312 (2009).

\bibitem{demanJVST} S. de Man, K. Haek, R. J. Wijngaarden and D. Iannuzzi, {\it J. Vac. Sci. Technol. B} {\bf 28}, C4A25 (2010).
\bibitem{Toricel} G. Torricelli, S. Thornton, C. Binns, I. Pirozhenko and A. Lambrecht, {\it J. Vac. Sci. Technol. B} {\bf 28}, C4A30 (2010).
\bibitem{ToricelEPL} G. Torricelli, I. Pirozhenko, S. Thornton, A. Lambrecht and C. Binns, {\it Europhys. Lett. } {\bf 93}, 51001 (2011).
\bibitem{Qun} Q. Wei, D. A. R. Dalvit, F. C. Lombardo, F. D. Mazzitelli and R. Onofrio, {\it Phys. Rev. A} {\bf 81}, 052115 (2010).
\bibitem{YalePRA} W. J. Kim, A. O. Sushkov,  D. A. R. Dalvit  and S. K. Lamoreaux, {\it Phys. Rev. A} {\bf 81}, 022505 (2010).
\bibitem{Ryan} R. O. Behunin, F. Intravaia, D. A. R. Dalvit, P. A. Maia Neto and S. Reynaud, {\it Phys. Rev. A} {\bf 85}, 012504 (2012).
\bibitem{Mos} H. T. Baytekin, A. Z. Patashinski, M. Branicki, B. Baytekin, S. Soh and B. A. Grzybowski, {\it Science} {\bf 333}, 308 (2011).
\bibitem{Joel} J. Laurent, H. Sellier, A. Mosset, S. Huant and J. Chevrier, {\it Phys. Rev. B} {\bf 85}, 035426  (2012).
\bibitem{Torg} S. K. Lamoreaux and J. R. Torgerson, {\it Phys. Rev. E} {\bf 71}, 036109 (2005).

\end{thebibliography}
\end{document}